\documentclass[twoside,a4paper,11pt]{sea10}
\usepackage{graphicx}
\usepackage{hyperref}
\usepackage{movie15}
\usepackage{float}
\begin{document}
\pagenumbering{arabic}
\pagestyle{myheadings}
\thispagestyle{empty}
{\flushleft\includegraphics[width=\textwidth,bb=58 650 590 680]{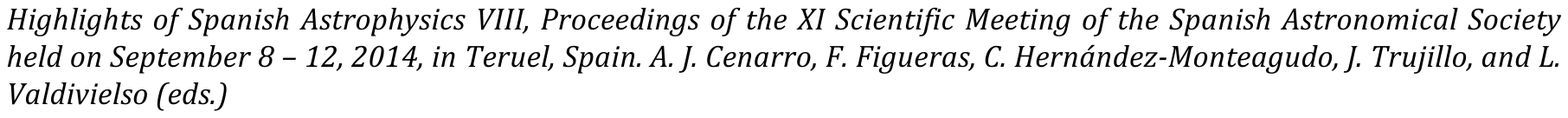}}
\vspace*{0.2cm}
\begin{flushleft}
{\bf {\LARGE
%
Present and future of the OTELO project
}\\
\vspace*{1cm}
%
M. Ram\'on-P\'erez$^{1,2}$,
A. Bongiovanni$^{1,2}$,
J. Cepa$^{1,2}$, 
A.M P\'erez-Garc\'ia$^{1,2}$,
E.J. Alfaro$^{3}$,
H. Casta\~neda$^{4}$
A. Ederoclite$^{5}$,
J.I. Gonz\'alez-Serrano$^{6}$,
J.J. Gonz\'alez$^{7}$,
J. Gallego$^{8}$
and
M. S\'anchez-Portal$^{9,10}$
%
}\\
\vspace*{0.5cm}
%
$^{1}$
Instituto de Astrof\'isica de Canarias, E-38200 La Laguna, Tenerife, Spain\\
$^{2}$
Departamento de Astrof\'isica, Universidad de La Laguna, E-38206 La Laguna, Tenerife,
Spain\\
$^{3}$
Instituto de Astrof\'isica de Andaluc\'ia (CSIC), Apdo. 3004, E-18080, Granada, Spain\\
$^{4}$
Instituto Polit\'ecnico Nacional, 07738 Distrito Federal, Mexico \\
$^{5}$
Centro de Estudios de F\'isica del Cosmos de Arag\'on, Teruel, Spain\\
$^{6}$
Instituto de F\'isica de Cantabria, CSIC-Universidad de Cantabria, Santander, Spain\\
$^{7}$
Instituto de Astronom\'ia, Universidad Nacional Aut\'onoma de Mexico, Apdo Postal 70-264,
Cd. Universitaria, 04510, Mexico\\
$^{8}$
Departamento de Astrof\'isica, Universidad Complutense de Madrid, 28040 Madrid, Spain\\
$^{9}$
European Space Astronomy Centre (ESAC)/ESA, P.O. Box 78, 28690, Villanueva de la Ca\~nada, Madrid, Spain\\
$^{10}$
ISDEFE, Madrid, Spain\\

%
\end{flushleft}
%
\markboth{
Present and future of the OTELO project
}{ 
%
M. Ram\'on-P\'erez et al.
%
}
\thispagestyle{empty}
\vspace*{0.4cm}
\begin{minipage}[l]{0.09\textwidth}
\ 
\end{minipage}
\begin{minipage}[r]{0.9\textwidth}
\vspace{1cm}
\section*{Abstract}{\small
%
OTELO is an emission-line object survey carried out with the red tunable filter of the instrument OSIRIS at the GTC, whose aim is to become the deepest emission-line object survey to date.  With 100\% of the data of the first pointing finally obtained in June 2014, we present here some aspects of the processing of the data and the very first results of the OTELO survey. We also explain the next steps to be followed in the near future.
%
\normalsize}
\end{minipage}
%
%
%
\section{Introduction \label{intro}}

Emission lines allow for the detection of very distant and faint objects that could otherwise remain invisibles to us. These lines can be found in all sorts of astronomical objects, such as nebulae or starburst galaxies, but are specially interesting and useful in extragalactic sources. Active galactic nuclei (AGN) and quasi-stellar objects (QSO) are among the objects that can be easily detected by means of their strong emission lines, in spite of being very distant. In the last years, great efforts have been made in order to develop new techniques and tools to detect emission lines. One of the most efficient ways is to use narrow-band photometry, which allows the simultaneous observation of all the sources within the field of view, and also a very deep exploration of the sky \cite{emissionline}. Tunable filters (TF) go one step further as they enable the tomographic sampling of the spectral range, thus making it possible to obtain 2D low resolution spectroscopy of all the objects in the field \cite{jones}. \par
OTELO (Osiris Tunable Emission Line Object) survey is the emission-line object survey carried out with the red tunable filter of the instrument OSIRIS at the GTC, the greatest fully steerable optical telescope in the world \cite{cepa2005}.  Four years after the beginning of the observations, and with more than 100 hours of time dedicated to them, we present in this contribution the very first results obtained with the 100\% of the data of OTELO's first pointing. We explain the techniques used in the reduction of the data and the extraction of the first preliminary catalogue, still to be improved.

\section{The OTELO survey\label{otelo_survey}}
The main goal of the OTELO survey is to study in detail different types of emitting objects and, by that, make a significant contribution to the field of extragalactic astrophysics. To this end, it targets all the objects that show emission lines in the field without any previous selection, so as to produce the biggest and deepest catalogue of emitting objects to date. The use of tunable filters makes it possible to obtain 2D low resolution spectroscopy of all the objects within the field of view, which is very extense (8 $\times$ 8 arcmin). This added to the large light-gathering capacity of the 10.4m telescope, allows a very deep observation of the sky. \par 

\begin{center}
\begin{table}[ht] 
\caption{EGS OTELO's observations} {\bigskip}
\center
\begin{minipage}{0.95\textwidth}
\center
\begin{tabular}{cccc} 
\hline\hline \noalign{\smallskip}
Year & Observation Block & Wavelength (\AA) & Mean seeing (arcsec)\\ [0.55ex]   
\hline \noalign{\smallskip}
2010 & 1--18 & 9280--9250 & 0.83 \\ 
2011 & 19--39 & 9244--9208 &  0.83 \\
2013 & 40--66 & 9202--9154 &  0.82 \\
2014 & 67--108 & 9148--9070 &  0.82 \\ \noalign{\smallskip}
\hline
\end{tabular} 
\end{minipage}
\label{tab1} 
\end{table}
\end{center}

OTELO focuses on the spectral range 9070-9280\AA$ $, matching a window in the airglow emission of the atmosphere. The technique used is the tomographic sample of the spectral range every 6\AA, with a full-width at half maximum of 12\AA. In this way, 36 frames with different central wavelengths are obtained, each one composed of 6 dithered images. The spectral resolution (R$\simeq$700) is set so as to deblend the H$\alpha$ and the [NII] lines \cite{lara2013}. OTELO expects to find different emission lines at different intervals of redshift: H$\alpha$ at $z\sim 0.4$, [OIII]5007 at $z\sim 0.83$, H$\beta$ at $z\sim 0.83$, or even Ly$\alpha$ at $z\sim 6.55$. The fields of view of OTELO are a region of the Extended Groth Strip (EGS) for the first pointing, and the central part of the Lockman Hole for the second one. In total, about 100 arcmin$^2$ of the sky will be covered. To this date, the observations of the first pointing, which started in 2010, have been completed (June 2014). More than 100 hours of observation with a mean seeing of 0.8 arcsec have been devoted to this task  (see Table \ref{tab1}). The observations of the second pointing will start in the Winter of 2014.\par


\section{Image reduction\label{reduction}}

Most of the reduction of the OTELO data was carried out using standard IRAF routines. The package TFRed, specially designed for the reduction of extragalactic images from tunable filters, was also used. First of all, the bias subtraction and trimming of the images were performed. The removal of cosmic rays was then completed with the L.A Cosmic task designed by van Dokkum \cite{vandokkum}. After that, the flat-field correction was carried out using a smooth \textit{superflat} resulting from the combination of dome flats' large scale and sky flats' small scale. Bad pixels in the images were also fixed by linear interpolation with bad pixel masks created for each different epoch.\par 
The following step in the reduction of the data was the sky rings subtraction, so as to obtain a background level of $\sim0$ counts in all images. As a consequence of the radial wavelength gradient which is characteristic of TF images\footnote{\label{note1} See the \href{https://docs.google.com/file/d/0B9hQ17il9GEoVWg0Ymx0RGFUcE92WTE2cUU3MWhKUQ/edit}{TF User Manual of OSIRIS}  at \href{http://gtc-osiris.blogspot.com.es/}{http://gtc-osiris.blogspot.com.es/} for more details.}, OH spectral bands emitted by the sky appear in the images as concentric circles (rings) around the optical center \cite{jones}. To remove them, a specific task called \texttt{ringsubV7} was created. This complex task first creates a mask of objects in the image to a certain sigma level and then evaluates the flux of the background in each one of these positions, by slightly rotating the image around the optical center and calculating the median value of the sky for each radius. The resulting image, that posseses the original background and, in the positions of the sources, the estimated one from the rotation procedure, is then smoothed with the \texttt{imsurfit} task. Finally, this model of the sky is subtracted to the original image. This tecnique is applied individually to each one of the images of OTELO.\par 
Once the ring subtraction is successfully applied, an additive correction is performed to remove the fringing pattern that arises as a consequence of interference effects due to multiple reflections of the light within the detectors (CCD). Monochromatic light, such as the one coming from intense spectral emission lines from the sky, is responsible for this effect. To remove it, images with the same central wavelength but different dithering are combined (after their sources have been masked), so as to create master fringe templates which are then subtracted to the original images with the \texttt{rmfringe} task.\par 
It is worth noting that even if the adoption of a dithering pattern for the 6 images of a slice is essential to remove the fringing effect and to subsequently identify \textit{ghosts}\textsuperscript{\ref{note1}} in the images that could be mistaken for real objects, it also implies that a given object will be located at slightly different distances from the optical center, that is, will be observed at a slightly different wavelenght in each of the images of a single frame. This effect is naturally taken into account when extracting the pseudo-spectra of the objects (see Section \ref{detection}).  \par 


\section{Astrometry and calibration\label{calibration}}

After the reduction process is completed, precise astrometry is performed in each individual image with a \textit{rms} not higher than 0.09'' (the scale of the images being $0.255$ arcsec/pixel). The images are then aligned in agreement with their newly provided coordinates and calibrated in flux and wavelength. First, the wavelength calibration is performed over the calibration stars so as to calculate the efficiency of the system. Two photometric standard stars (one in each CCD) that have been previously observed with OSIRIS long-slit spectroscopy over the same wavelength range as OTELO \cite{cesare} are used for this purpose. For each individual image, the wavelength calibration requires the exact position of the optical center and the central wavelength to which the TF was tuned to be known. The wavelength gradient is then taken into account by means of an empirical equation which relates the effective observation wavelength of a given source in each slice, $\lambda$, to the one the TF was tuned to ($\lambda_c$) and the radial distance to the optical center ($r$) \cite{gonzalez}:

\begin{displaymath} 
\lambda= \lambda _c  - 5.04r^2 + a_3(\lambda)r^3  
\end{displaymath}
\begin{displaymath}
a_3(\lambda) = 6.17808 -  1.6024 \times 10^3 \lambda + 1.0215 \times 10^7 \lambda ^2 \end{displaymath}

The flux of a given source  in each slice is then converted from ADU (Analog-to-Digital Unit) to physical units (erg/s/cm$^2$) with the following relation, that takes into account the gain of the CCD ($g$), the extinction ($K$, calculated with the mean values of the extinction coefficient and the airmass during the night), the energy of the photon ($E_\gamma$), the exposition time ($t$), the telescope's aperture ($A_{tel}$) and the efficiency of the system ($\epsilon$) \cite{jones}:

\begin{displaymath}
F = \frac{gK(\lambda)E_\gamma}{t.A_{tel}. \epsilon}  F_{ADU}
\end{displaymath}

\section{Detection and extraction of the sources\label{detection}}

The photometric analysis of the sources is made with the software SExtractor in dual mode. A \textit{deep} image, shown in Fig. \ref{pseudo}, is constructed by combining the 36 frames of the tomography, in order to obtain a full sensitivity image of the field to be used for source detection. A 3$\sigma$ level over the background noise is imposed as the minimal flux for a source to be detected, which gives a limiting flux of $\sim 1.04\times 10^{−-19}$ erg/cm$^2$/s. Once the positions of all the sources have been recorded with high precision, the next step is to extract their fluxes in each one of the 36 frames. The mean flux of a particular object in the 6 dithered images of a single frame of the tomography is recorded, as well as its mean observation wavelength in that frame. \par 
Doing so for each one of the 36 frames of the tomography, one can then construct the pseudo-spectrum of the object, where each point represents its integrated flux in a frame, that is to say, the result of the convolution between its spectrum and the TF response. The pseudo-spectra of all the objects found in the field above 3$\sigma$ with respect to the pseudo-continuum are then analysed with automatic algorithms so as to detect peaks and select those which could be showing an emission line. The pseudo-spectra of those emitting candidates are then visually analysed in a second round so as to discard false detections corresponding to cosmic rays residuals.

\section{First results\label{results}}

\begin{figure}[H]
\center
\includegraphics[scale=0.3]{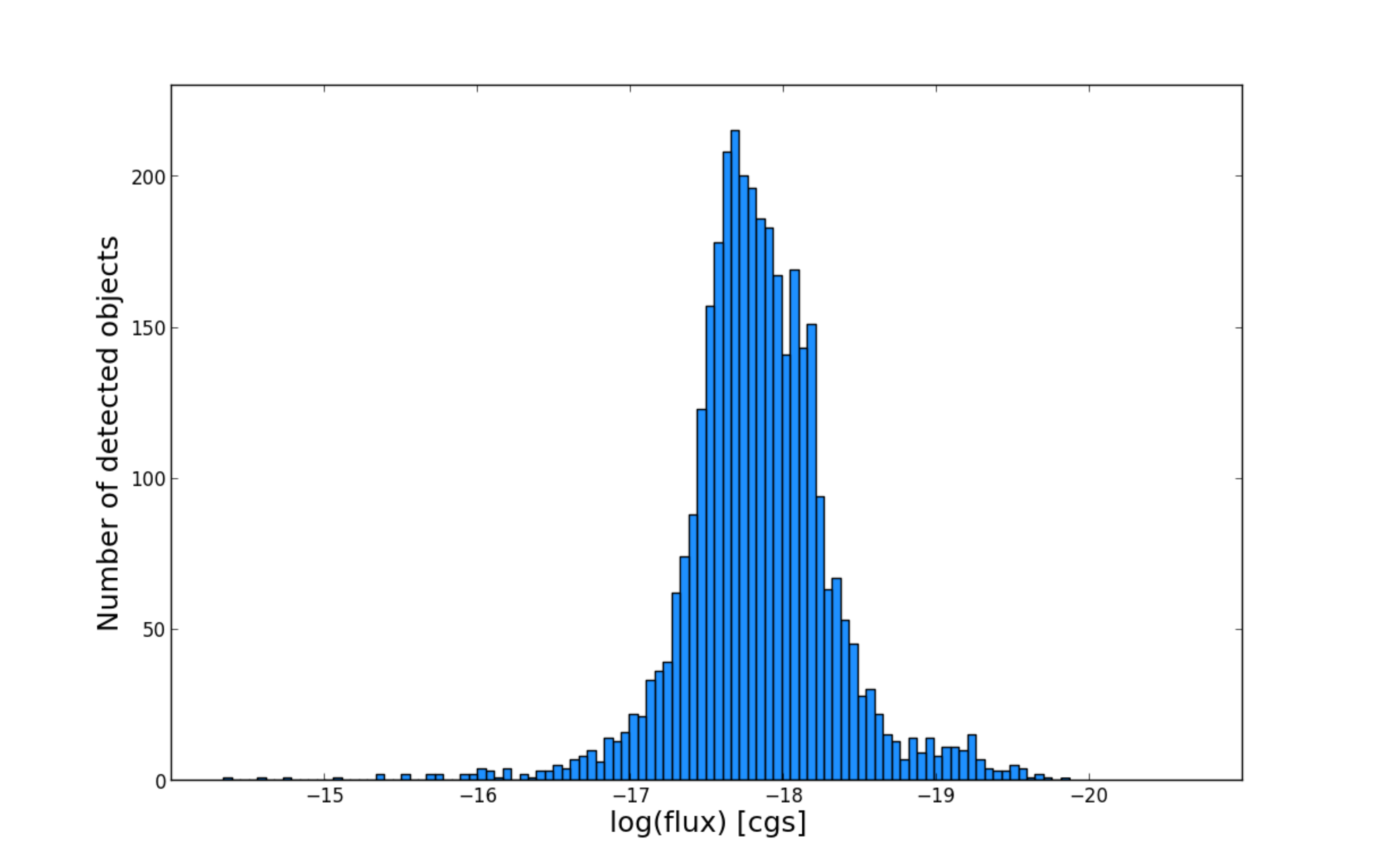} ~
\caption{\label{hist} Histogram of the 3472 sources detected in the EGS field of OTELO. A minimum detectable flux of $\sim 1.04\times 10^{-−19}$ erg/cm$^2$/s at 3$\sigma$ level was obtained.
}
\end{figure}

In a first analysis, 3472 sources were detected at 3$\sigma$ level over the deep image of OTELO survey in the EGS field. Figure \ref{hist} shows the histogram of all the detected sources. From it, a completitude flux, i.e, the flux up to which there is a complete sample of objects, was estimated at $\sim 2\times 10^{−-18}$ erg/cm$^2$/s. Also, a limiting flux of $\sim 1.04\times 10^{−-19}$ erg/cm$^2$/s at 3$\sigma$ level was calculated. This flux exceeds the one that was expected for OTELO \cite{cepa2011}, and also the one reached by other analogous surveys such as the Subaru Deep Field Survey \cite{subaru}, whose NB921 filter is comparable to the spectral window of OTELO.\par

\begin{figure}[H]
\center
\includegraphics[scale=0.41]{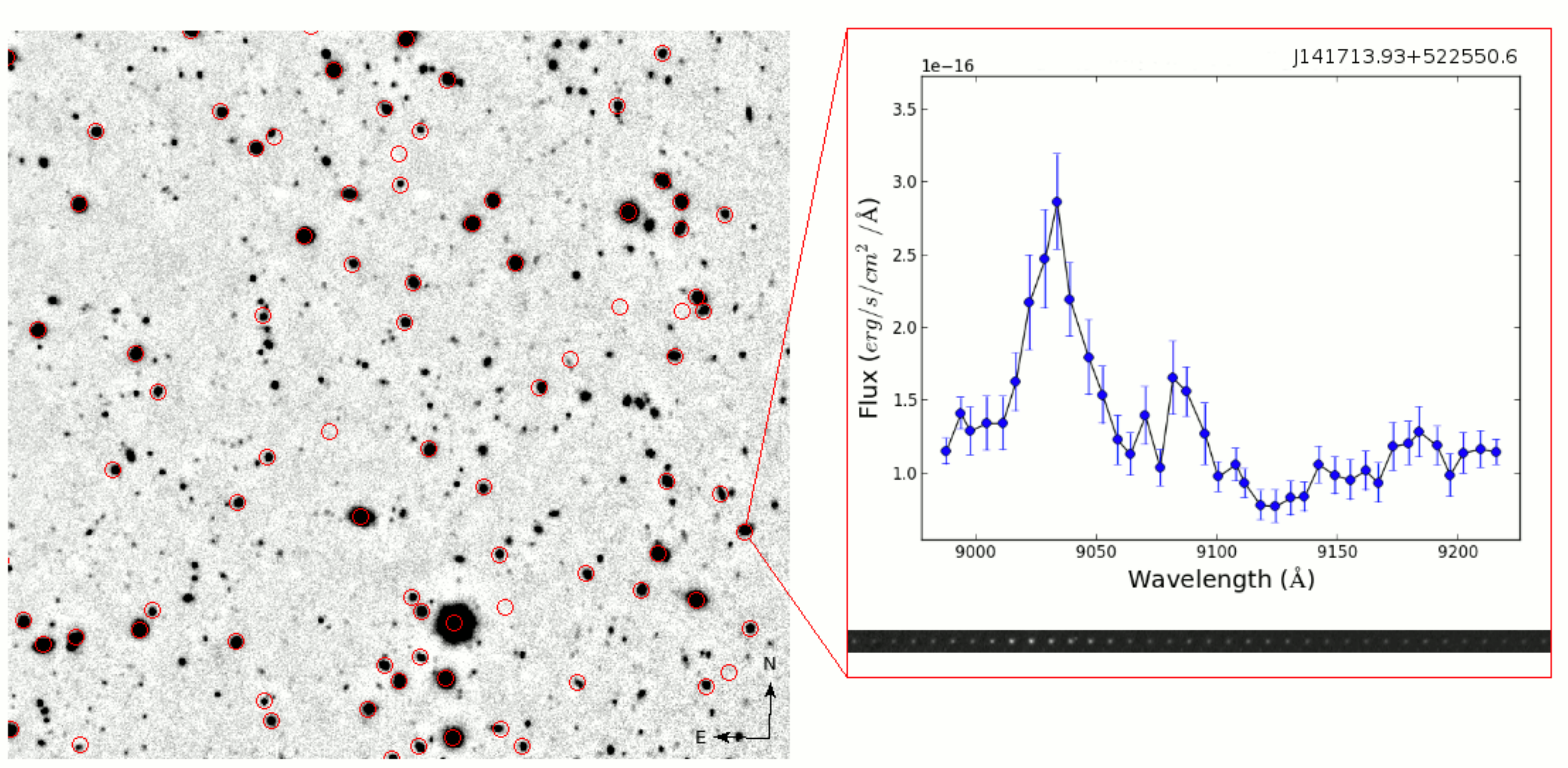} ~
\caption{\label{pseudo}  SW quadrant (3 x 3 arcmin$^2$) of the OTELO deep image (limiting flux $\sim 1.04\times 10^{−-19}$ erg/cm$^2$/s, 3$\sigma$). Red circles represent SDSS-DR9 photometric sources in the field. In the right side, a typical pseudo-spectrum of a confirmed H$\alpha$+[NII] emitter (J141713.93+522550.6; $z = 0.375$) as seen by OTELO. The strip below the plot is composed by the 36 slices of the corresponding object datacube.
}
\end{figure}

From all the objects detected in the fied, at least 10\% presented emission features, once the most obvious false detections, such as cosmic rays residuals, were discarded. An example can be seen in Fig. \ref{pseudo}, where the pseudo-spectrum of an object with an evident emission line is shown. This figure also displays a strip with the source's appearance in each one of the frames of the tomography. A variation in its intensity, corresponding to the emission of a spectral line, can clearly be seen.

\section{Conclusions and future work\label{summary}}


Although the results presented in this contribution are not conclusive as they need to be further refined, the preliminary analysis of the 100\% of the data of OTELO's first pointing appears to be promising. The depth and sensitivity attained with the observations are very high when compared to those of its closest competitors: the narrow-band survey of Subaru at $\sim$920nm, for instance, has a limiting flux of $\sim 2\times 10^{−-18}$ erg/cm$^2$/s at 3$\sigma$ and an equivalent-width (EW) of observation of 132\AA$ $ \cite{subaru}. As seen in this contribution, OTELO's data go far beyond, both in limiting flux (almost 20 times deeper) and in EW (one order of magnitude more sensitive). The pseudo-spectra that have been extracted also present very good signal-to-noise ratios so as to detect emission lines. All of this makes OTELO an unique survey to detect new distant objects with emission lines.\par 
The next step will be the refinement of the procedure with better algorithms so as to automatically discard false detections (cosmic rays residuals, \textit{ghosts}...), as previously mentioned in Section \ref{detection}. Then, the catalogue will be expanded with complementary data from other surveys (Chandra, Galex, CFHT-LS, Herschel, Spitzer, AEGIS...) over the whole spectral range (X-rays, UV, optical, IR). This will allow to obtain the spectral energy distribution of the sources with which the photometric redshift will be estimated. Once $z$ is known, it will be straightforward to identify the chemical species responsible for the spectral lines. Their fluxes and equivalent widths will then be extracted by modelling the pseudo-spectra. If necessary, follow-up spectroscopy will be performed for the most interesting objects.\par 
One of the lines of research in which the data of OTELO will be significant is the study of galaxies with active galactic nuclei (AGN). OTELO's catalogue will allow to distinguish AGN from starburst-galaxies by means, for example, of optical diagnostic diagrams such as the one proposed by Cid-Fernandes et al (2010) \cite{cid}, color-color diagrams, or with the help of the complementary data, like the X-ray catalogue of Povi\'c et al (2009) \cite{povic}.

\small  
%
\section*{Acknowledgments}   
This work was supported by the Spanish Ministry of Economy and Competitiveness (MINECO) under the grant AYA2011-–29517-–C03–-01. Based on observations made with the Gran Telescopio de Canarias (GTC), installed in the Spanish Observatorio del Roque de los Muchachos of the Instituto de Astrofísica de Canarias, in the island of La Palma.  
%

%
\end{document}